\documentclass{article}

\usepackage{blindtext}
\usepackage{amsmath}
\usepackage{graphicx}
\usepackage{subcaption}
\usepackage{float}
\usepackage[ruled,vlined]{algorithm2e}
\usepackage{hyperref}

\usepackage{booktabs}
\usepackage{array}
\usepackage{longtable}
\usepackage{siunitx} 
\usepackage{multirow}

\usepackage{authblk} 

\title{A Bio-Inspired Minimal Model for Non-Stationary K-Armed Bandits}

\author[1]{Krubeal Danieli}
\author[2]{Mikkel Elle Lepperød}
\affil[1]{Center for Integrative Neuroplasticity, FYSCELL, University of Oslo, Norway}
\affil[2]{Simula Research Laboratory, Oslo, Norway}

\date{} 

\begin{document}

\maketitle

\begin{abstract}
While reinforcement learning algorithms have made significant progress in solving multi-armed bandit problems, they often lack biological plausibility in architecture and dynamics.
Here, we propose a bio-inspired neural model based on interacting populations of rate neurons, drawing inspiration from the orbitofrontal cortex and anterior cingulate cortex.
Our model reports robust performance across various stochastic bandit problems, matching the effectiveness of standard algorithms such as Thompson Sampling and UCB.
Notably, the model exhibits adaptive behavior: employing greedy strategies in low-uncertainty situations while increasing exploratory behavior as uncertainty rises. Through evolutionary optimization, the model's hyperparameters converged to values that align with known synaptic mechanisms,
particularly in terms of synapse-dependent neural activity and learning rate adaptation.
These findings suggest that biologically-inspired computational architectures can achieve competitive performance while providing insights into neural mechanisms of decision-making under uncertainty.
\end{abstract}



\hfill \break
\vspace {0.5cm}

\section{Introduction}

The ability to make decisions for long-term reward maximization is a fundamental aspect of cognition. The brain has evolved specialized and interconnected regions to implement this behaviour under the constraints of biology.

Well-studied ecological settings for decision making are foraging tasks, such as food search. In these problems, the agent is usually asked to choose between different options to maximize an expected reward.
In nature, animals have been shown to exhibit different strategies depending on context.
\textit{Matching behaviour} is a well-known phenomenon in which the animal's decision patterns are proportional to the reward probability of the available options.
This behavior is believed to be the result of the trade-off between exploration and exploitation \cite{suttonReinforcementLearningProblem1998, nivEvolutionReinforcementLearning2002}.
In fact, this is a well known phenomenon in the reinforcement learning literature, in which an agent is faced with the dilemma of exploring new alternatives, potentially more rewarding, or exploiting known options, despite being possibly sup-optimal.

A popular formalization of these types of tasks is the \textit{multi-armed bandit} problem (MAB) \cite{averbeckTheoryChoiceBandit2015}. This setting is usually described in terms of a slot machine endowed with $K$ distinct arms, also called levers.
During a round, the agent selects one of the arms and collects a reward $R$ according to an unknown probability of reward specific to the chosen arm.
The goal is simply to maximize the total reward after a given number of steps, which is achieved by effectively updating a selection policy after each round.
This problem has been extensively studied in the context of reinforcement learning and is considered a fundamental building block for more complex tasks \cite{suttonReinforcementLearningProblem1998}.


The multi-armed bandit problem comes in several variants, with the simplest featuring a stationary reward distribution.
An important performance measure in these tasks is \textit{regret}, usually defined as the distance between the selected choice and the theoretically optimal one.
Researchers have proposed numerous algorithms to address this problem, each with distinct theoretical guarantees.

Thompson sampling (TS) is a widely adopted approach rooted in Bayesian optimization. It maintains a posterior distribution over action reward probabilities and selects actions by sampling from these distributions. Thompson sampling has demonstrated near-optimal regret bounds in stochastic settings \cite{agrawalAnalysisThompsonSampling2012, kaufmannThompsonSamplingAsymptotically2012}.

In contrast, the Upper Confidence Bound (UCB) algorithm uses an optimistic principle for exploration. It maintains an estimate of the reward for each option by a confidence interval. Action selection relies on the upper bound of this interval, encouraging exploration of less-visited options by assigning them higher uncertainty. UCB has been shown to achieve logarithmic regret in stochastic bandits \cite{auerFinitetimeAnalysisMultiarmed2002}.

Another effective baseline is the $\epsilon$-Greedy strategy. At each decision step, a random action with probability $\epsilon$ and the best known action with probability $1 - \epsilon$ (exploitation) is selected.
Although not as theoretically optimal as Thompson Sampling or UCB, $\epsilon$-Greedy is simple to implement and often effective in practice.
Extensions such as VDBE adapt $\epsilon$ dynamically based on the variance of the value function, providing better control over exploration \cite{gittinsBanditProcessesDynamic1979, banMultifacetContextualBandits2021, tokicAdaptiveEGreedyExploration2010, tokicValueDifferenceBasedExploration2011}.

However, these traditional algorithms, despite their effectiveness, lack biological plausibility – they neither resemble neural circuits nor follow synaptic plasticity dynamics.
For example, they do not rely on a network-like architecture with interconnected units, as seen in the brain.
Additionally, their action selection process is typically instantaneous, whereas decision making in the brain occurs over time, often requiring the activity of a neural circuit to evolve and stabilize before converging on a final selection.

Their learning mechanisms also differ fundamentally. They typically involve explicit updates to statistical parameters (e.g., reward estimates or exploration rates) based on observed outcomes.
In contrast, biological learning relies on local plasticity rules, where synaptic changes depend on the activity of connected neurons, modulating how input is integrated and how output signals are generated.

Although not the primary driver, these limitations align with a growing interest in machine learning towards bioinspired algorithms, such as neural networks and predictive coding \cite{lee2022, spratlingReviewPredictiveCoding2017}, offering several advantages.

In fact, these methods can achieve state-of-the-art performance in various domains, including the challenging \textit{machine-challenging tasks} (MCTs), set of problems that are difficult for machines but relatively easy for humans \cite{schmidgallBraininspiredLearningArtificial2024, hassabisNeuroscienceInspiredArtificialIntelligence2017, lee2022}.
In addition, bioinspired models enhance algorithmic interpretability by clarifying the functional relationships between internal components.
When applied to tasks with existing experimental data, these models can generate new insights into the brain and suggest new research directions \cite{liuSeeingBelievingBrainInspired2023}.
Although other approaches such as Bayesian learning can demonstrate optimal performance and match human data well \cite{behrens2007}, they are more difficult to relate to neuronal dynamics.

\hfill \break

\indent In this work, we aim to enhance the biological plausibility of models used in multi-armed bandit tasks by introducing a novel, minimal decision-making architecture called Neural Selection Agreement model (NSA).
This model comprises two interacting rate-based neuronal populations connected by plastic synapses, uses a biologically inspired plasticity rule, and forms decisions based on the agreement of the two populations on the next option.


The model's plasticity mechanism is non-Hebbian and depends on the magnitude of inter-population synaptic weights.
This formulation aligns with synapse-type specific plasticity (STSP), a biologically supported mechanism linking learning dynamics to synaptic resource availability, current state, and morphological properties \cite{larsenSynapsetypespecificPlasticityLocal2015, blackmanTargetcellspecificShorttermPlasticity2013, bartolHippocampalSpineHead2015, arielIntrinsicVariabilityPv2012}.
Learning rates are also adaptive, in line with observations in human experiments \cite{behrens2007}.

Similar forms of plasticity have been employed in prior work on spiking neural networks and models of synaptic metaplasticity \cite{inglis2021, iigaya2016}.
Despite its simplicity, our model performs comparably with standard algorithms such as Thompson Sampling, $\epsilon$-Greedy and Upper Confidence Bound, while offering a more neurobiologically grounded account of decision-making.

Other studies have also proposed solutions to bridge reinforcement learning and neural mechanisms.
For example, \cite{starkweather2018} proposed temporal difference algorithms of the belief state \cite{babayan2018} that abstract dopaminergic signaling and medial prefrontal projections.
These models highlight the role of hidden-state inference during probabilistic tasks, although they assume fixed reward distributions and do not incorporate explicit synaptic plasticity. 

In \cite{khorsand2017}, metaplasticity mechanisms were explored in relation to the probability estimation of binary sequences, uncovering informative patterns of functional synaptic states.
However, the environment varied along a single stimulus dimension.
Similarly, \cite{farashahi2017} applied a metaplasticity model to a probabilistic reversal learning task, effectively a stationary two-armed bandit, revealing the emergence of option-specific learning dynamics.

A notable exception is \cite{iigaya2016}, which addressed a non-stationary two-arm bandit using a synaptic cascade model \cite{fusi2005} equipped with a surprise detection mechanism to track changes in reward probability.

In contrast to previous work, our study addresses more challenging, high-dimensional nonstationary reward environments, including up to 1,000 arms with independently drifting reward probabilities.
We specifically focus on stochastic bandit problems with \textit{concept drift}, where reward distributions evolve over time, either gradually or through abrupt changes, thus requiring flexible and adaptive decision-making strategies \cite{garivierUpperConfidenceBoundPolicies2008, besbesStochasticMultiArmedBanditProblem2014, cavenaghiNonStationaryMultiArmed2021}.

Despite its simplicity, the proposed model performs competitively with standard algorithms such as Thompson Sampling, $\epsilon$-Greedy, and Upper Confidence Bound, while offering a more biologically grounded perspective on decision-making mechanisms.

In general, our work aims to bridge adaptive decision-making under uncertainty with principles from computational neuroscience.
By proposing a biologically plausible mechanism for choice behavior in dynamic environments, we contribute a framework that may inform both the development of adaptive artificial systems and the interpretation of neural processes underlying flexible behavior.

\hfill \break
The remainder of this paper first describes our model design and learning, then presents experimental results and comparative analyses with established algorithms, and lastly discusses the findings' broader implications and potential future directions.




\section{Methods}

\noindent The following section is organized as follows. First, we introduce a formalization of the general problem setting, together with the variants considered in this work. Then we outline the architecture of our model and how it can be mapped to neurobiology. Finally, we describe the learning procedure
and showcase its dynamics in a simple example.

\subsection{Binomial MAB problem}
\hfill \break
\noindent The standard formulation of the task is structured as a set of $K$ arms (or levers) $\mathcal{A}_{K}=\{a_{1}\ldots a_{K}\}$, with an associated reward distribution $\mathbf{p}=\{p_{1}, \ldots p_{K}\}$.
At each iteration, the agent pulls an arm and collects a possible reward drawn as a Bernoulli variable $R\sim \mathcal{B}(\{0,1\},p_{k})$. The agent's objective is to maximize the total reward $\sum^{T}_{t} R_{t}$, after a certain number of rounds $T$, also called the horizon.
Importantly, the agent is unaware of the true probability of reward and therefore has to make its decisions following a certain policy, denoted $\pi$.
In the reinforcement learning literature, the policy is often defined as a distribution over actions, here the arms $\mathcal{A}_{K}$, given the current state at time $t$. In the bandit problem, the state can be taken to correspond to the history $h_{t}$ of past actions and rewards in the period
$(0\ldots t]$, and the policy as a function that returns a selected arm $\pi(h_{t})=a_{t}$ \cite{qiForcedExplorationBandit2023}.

Given the inherent stochasticity of the feedbacks from the environment, the policy is affected by the so-called exploration-exploitation trade-off, which here is phrased as the contrast between the option of the arm with the estimated highest expected reward versus the option to explore other arms, so as to gather more information.
A common approach is the $\epsilon$-Greedy policy, where the choice to explore is selected with probability $\epsilon$.
Moreover, it is often preferable to have more explorative behavior early during the training, with the intent to have a good sample size for the empirical reward distribution, which can be later exploited for maximizing reward.

Another important concept in multi-armed bandit problems is \textit{regret}. Intuitively, it quantifies the loss of reward due to following a certain policy, and it is determined by the difference between the collected reward and the theoretical optimal, obtained by choosing the best arm at each round.
Formally, given defined a function $r(\pi)$ that returns the expected reward while following policy $\pi$, the regret $\rho$ over an horizon $T$ can be formulated as:
\begin{equation}
    \rho = \frac{1}{T}\sum^{T}_{t} p^{*}_{t} -r(\pi(h_{t}))
\end{equation}

\noindent where $p^{*}_{t}$ is the expected reward of the optimal arm at time $t$, which corresponds to its probability since it is a Bernoulli distribution.
\noindent The goal of the agent is to minimize the regret and thus maximize the total reward.


\subsection{Neural Selection Agreement model (NSA)}
The model is constructed as a rate network composed of two neuronal populations, \textit{U} and \textit{V}.
The first, \textit{U}, represents the memory traces of the \textit{K} available options (\textit{ that is,}, the bandits), while the second, \textit{V}, encodes their values according to current policy.

In our model, the first layer represents the available options, while the learned connections to the second layer encode their values based on recent reward history. A key simplification is the lumping of option representations into single neurons. Although this choice abstracts the more distributed encoding found in actual brain networks, it allows for a more tractable model design \cite{martinRepresentationObjectConcepts2007a}.

More formally, the model is defined by a set of coupled ordinary differential equations (ODE). The first equation describes the evolution of neural activity $\textbf{u}$ in population \textit{U}, while the second governs the activity $\textbf{v}$ in population \textit{V}, each evolving with its respective time constant $\tau$.

\begin{equation}
\begin{aligned}
    \tau_{u} \dot{\textbf{u}}&= -\textbf{u} + \textbf{W}^{VU}\phi_{v}(\textbf{v}) + \textbf{I}_{\text{ext}} \\
    \tau_{v} \dot{\textbf{v}}&= -\textbf{v} + \widetilde{\textbf{W}}^{UV}\phi_{u}(\textbf{u})
\end{aligned}
\end{equation}\label{eq:main}

\noindent The external input $\textbf{I}_{\text{ext}}$ is a constant input that is used to set the initial conditions of neural activity $\textbf{u}$.
The activation functions $\phi_{v},\phi_{u}$ are applied to population $v$ and $u$ respectively, and represent two distinct neural response functions tailored to each population vector.
They have been chosen to be a step function with threshold $\theta_{v},\theta_{u}$ applied to a generalized sigmoid with gain $g_{v},g_{u}$ and offset $s_{v},s_{u}$.

\begin{figure}[ht]
    \centering
    \includegraphics[width=0.6\textwidth]{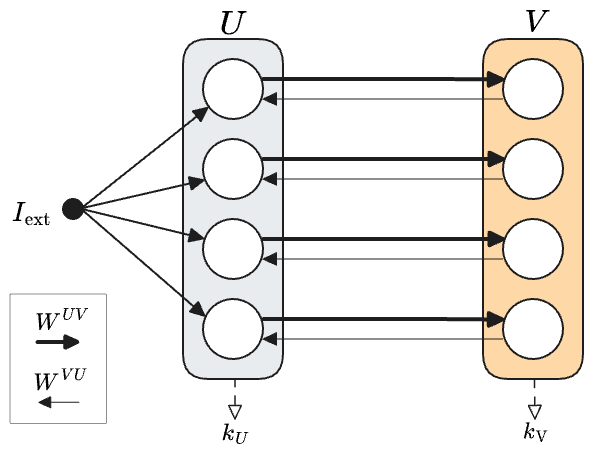}
    \caption{\textsc{Model architecture} - \textit{The model is composed of a layer $U$ (grey), receiving a feedfoward input $I_{\text{ext}}$, a layer $V$ (orange), and connections $\textbf{W}^{UV}$ and $\textbf{W}^{VU}$. Additionally, two indexes $k_{U}, k_{V}$  are extracted from the layers and
    corresponds to the selection made by the two populations as $k_{U}=\text{argmax}_{k} \{\textbf{u}\}$, $k_{V}=\text{argmax}_{k} \{\textbf{v}\}$.}}
    \label{fig:main_architecture}
\end{figure}

\noindent Importantly, the two layers are not fully connected and the matrices are diagonal.
More in detail, the weight matrix $\textbf{W}^{VU}$ is simply made of $1$s, while $\widetilde{\textbf{W}}^{UV}$ is a function of the actual weights $\Phi_{v}(\textbf{W}^{UV})$ and represents the contribution of the active options $\textbf{u}$ to the value representation $\textbf{v}$, so it is called \textit{ the option value function}. The matrix $\textbf{W}^{UV}$ is initialized to all zeros.
The function $\Phi_{v}$ is defined as the weighted sum of a generalized sigmoid and a Gaussian, whose shape is characterized by a bell curve that settles smoothly to a constant value. For details, see the Appendix \ref{sec:appendix}.

The motivation behind our choice of $\Phi_{v}$ is to be agnostic about its final form and to allow competition or integration of two distinct characteristics of the shape of the function.
In particular, it corresponds to a smooth transition to a plateau value with a certain steepness (or gain), which can represent a saturation once a threshold is crossed; such features have been reported for biological and artificial neurons \cite{ockerFlexibleNeuralConnectivity2020, apicellaSurveyModernTrainable2021}.
The other is a bell-shaped curve with a defined center and width, which can allow putting emphasis on values only within a given window and modulating information transfer \cite{millerCombinedMechanismsNeural2019}.

The model hyperparameters were optimized to maximize the average total reward over multiple runs. In particular, given the non-differentiability of the model with respect to the fitness function, we employed an evolutionary algorithm, more specifically CMA-ES.

\subsubsection{Option selection}
The decision-making process within a single round is structured in two distinct phases. Initially, the model receives a constant external input that targets all neurons in the memory population \textit{U} equally.
During this phase, $\textbf{I}_{\text{ext}}$ works as an equilibrium value while reciprocal interactions with population \textit{V} push $\textbf{u}$ to different values, depending on the current policy encoded in $\widetilde{\textbf{W}}^{UV}$.
Importantly, the weights $\textbf{W}^{UV}$ are initialized to zero, and thus the input from $U$ to $V$ is uniform. This approach ensures the absence of biases towards any arm by having all weights equal, and corresponds to a completely untrained network.
After a fixed time $\sim 2 \text{s}$, the second phase begins. Here, the external input is removed and the model is left to evolve autonomously, and since there are no recurrent connections in neither population, the dynamics are entirely driven by their coupling.
A selection $k$ is sampled after another fixed amount of time $\sim 5 \text{s}$, and is defined according to the following rule:

\begin{equation*}
    k =
    \left\{
        \begin{array}{ll}
            \text{argmax}_{k}\{\textbf{v}\} & \text{\textit{if}}\; \text{argmax}_{k} \{\textbf{v}\} = \text{argmax}_{k} \{\textbf{u}\} \\
            \text{random}(K) & \text{\textit{otherwise}}
        \end{array}
    \right.
\end{equation*}

\noindent The selection rule is simple: if the value representation $\textbf{v}$ is in agreement with the memory trace $\textbf{u}$, then the option with the highest value is selected. Otherwise, a random option is chosen.
This rule presents a way to express the exploration-exploitation trade-off through the possible agreement of the two populations, under the influence of current weight values $\widetilde{\textbf{W}}^{UV}$. \\ In subsection \ref{alg:decision1}, the pseudo-code for the algorithm behind the selection process is reported below, which is applied during each round $t$.

\begin{algorithm}[ht]
\caption{Two-phases option selection process}
\label{alg:decision}
\SetAlgoLined
\KwIn{External input $\textbf{I}_{\text{ext}}$, population $\textbf{u}$, population $\textbf{v}$, weights $\widetilde{\textbf{W}}^{UV}$}
\KwOut{Selected action $k$}

\SetKwComment{Comment}{// }{ }

\textbf{Phase 1:} \textit{external input} \Comment*[r]{Duration: $\sim$2s}
Define constant $\textbf{I}_{\text{ext}}$\;
Update populations $\textbf{u}, \textbf{v}$ according to \ref{eq:main}\;

\textbf{Phase 2:} \textit{autonomous evolution} \Comment*[r]{Duration: $\sim$2s}
Remove external input $\textbf{I}_{\text{ext}}$\;
Let system evolve through population coupling according to \ref{eq:main}\;

\textbf{Selection process:}\;
$k_{u} \gets \text{argmax}_{k}\{\textbf{u}\}$\;
$k_{v} \gets \text{argmax}_{k}\{\textbf{v}\}$\;
\eIf{$k_{u} = k_{v}$}{
    $k \gets k_{v}$ \Comment*[r]{Exploitation}}{
    $k \gets \text{random}(K)$ \Comment*[r]{Exploration}
}
\Return $k$
\end{algorithm}\label{alg:decision1}

\noindent Lastly, the structure of the option selection process resembles the prefrontal circuitry, as the choices emerge from the state sampling of the network following a period of autonomous neural activity. The stability of these neural activations depends on the strength and reliability of the option with the highest value \cite{backmanEffectsWorkingMemoryTraining2011, enelStableDynamicRepresentations2020}.

\noindent According to the values of the policy parameters, the behavior of the model displays periods of exploration followed by steady exploitation, which can be reverted in the case of a change in the environment's reward distribution.

\subsection{Learning}
Given a selected option $k$, the environment (set of bandits) samples and returns a reward $R\in \{0, 1\}$ with probability $p_{k}$.
Then, the weights $\textbf{W}^{UV}$ for the neuron corresponding to option $k$ are updated according to the following plasticity rule.

\begin{equation}
    \Delta \textbf{W}^{UV}_{k} = \tilde{\eta}_{k} \left(R\cdot w^{+}- \textbf{W}^{UV}_{k}\right)
\end{equation}

\noindent where $w^{+}$ is a constant maximum synaptic weight, while $\tilde{\eta}_{k}$ is the learning rate for the option $k$ determined by a function $\Phi_{\eta}$ of the current weights $\textbf{W}^{UV}_{k}$, known as the \textit{learning rate function}.

The shape of $\Phi_{\eta}$ is again a Gaussian-sigmoid but with different parameters, giving evolution the opportunity to combine the two characteristic traits of the plateau and the bell-shaped tuning.
In particular, these characteristics can be combined to define mechanisms of synapse-type specific plasticity as a function of current synaptic strength \cite{larsenSynapsetypespecificPlasticityLocal2015}, as well as the application of other useful homeostatic constraints with computational advantages, such as synaptic scaling and proportional updates \cite{citriSynapticPlasticityMultiple2008, kennedySynapticSignalingLearning2016, samavatSynapticInformationStorage2024}.





\section{Experiments}

The NSA model has been tested in a series of benchmark environments, each with a different number of arms and reward distributions. The performance has been compared with the following algorithms: Random Baseline, Upper-Confidence Bound (UCB), Thompson Sampling, and Epsilon-Greedy.

\subsection{Game variants}\label{sec:envs}

\begin{figure}[ht]
    \centering
    \includegraphics[width=1.\textwidth]{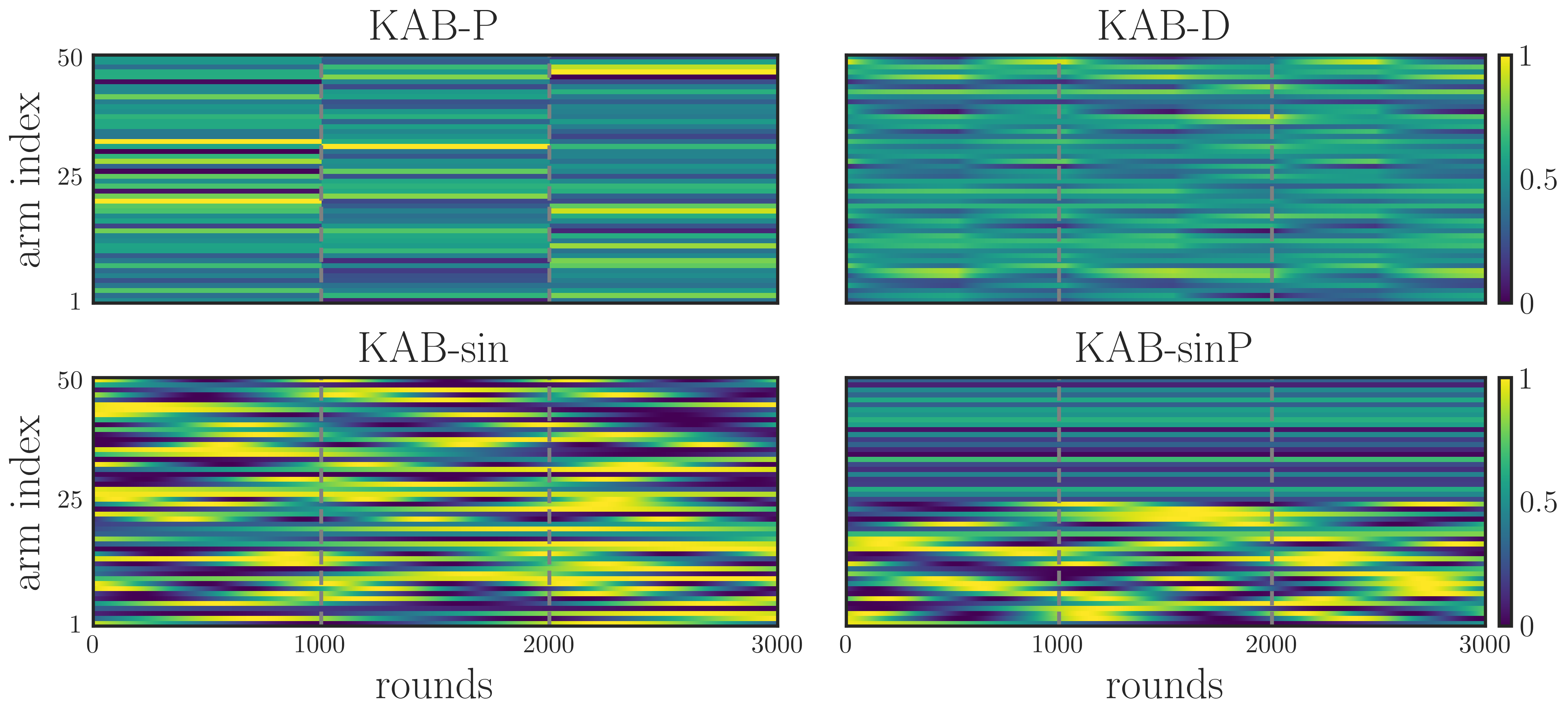}
    \caption{\textsc{Reward distribution for the four MAB variants} - \textit{The reward distribution for each variant is illustrated: piecewise stationary distribution (MAB-P), piecewise stationary distribution with drift (MAB-D), sinusoidal distribution shift (MAB-sin), partial sinusoidal distribution shift (KAB-sinP).
    The arms are organized in rows; the grey dashed lines demarcate trials (3), which are block of rounds represented by columns, and here only reporting the arm reward proability.}}
    \label{fig:envs}
\end{figure}

\noindent Our goal is to investigate the performance of the agent in a non-stationary environment, meaning that its underlying distribution changes over time \footnote{Since the arm probabilities are not normalized to $1$, it is
technically improper to call them \textit{probability distributions}; we will therefore refer to either \textit{probability} or \textit{distribution} separately at any given time for avoiding confusion.}
We choose this setting as it resembles an scenario in which an animal has to forage in an environment with food (reward) distributed over a set of fixed locations, but whose occurrence probability can change over time.
A \textit{round} -or horizon- is defined as an action-reward event; instead, a \textit{trial} is a a block of rounds.
For testing, four slightly different MAB variants were used, obtained by introducing different types of non-stationarity: piecewise constant, uniformly changing, sinusoidally changing, and sinusoidally changing with piecewise constant arms.
The reason for these choices is to test the model performance under different speed and uniformity of the distribution changes.
Figure \ref{fig:envs} visually illustrates their specificities.

\hfill \break
\noindent \textbf{Piecewise stationary distribution} [\textsc{MAB-P}]\\
Within a trial the reward distribution is stationary and it is drawn from a normal $\mathbf{p}=\mathcal{N}(0.5, 0.2)^{K}$, clipped in $(0, 1)$. At the end of each trial $i$ it is drawn a new distribution $\mathbf{\mathbf{p}}_{i} \to \mathbf{\mathbf{p}}_{i+1}$ \cite{qiForcedExplorationBandit2023}.

\hfill \break
\noindent \textbf{Piecewise stationary distribution with drift} [\textsc{MAB-D}]\\
At the very beginning, the reward distribution $\mathbf{p}$ is sampled from a normal $\mathbf{p}=\mathcal{N}(0.5, 0.2)^{K}$.
Then, it changes gradually over the rounds, tracked as time $t$, such that its values tend towards a target distribution $\mathbf{q}_{i}$ as $\tau_{p}\dot{\mathbf{p}}_{t}=\mathbf{q}_{i}-\mathbf{p}_{t}$.
Here, $\dot{\mathbf{p}}$ is the time derivative of the distribution and $\tau_{p}$ is its time constant.
Once the distance is below a threshold $\delta$ as $\vert \mathbf{q}_{i} - \mathbf{p}_{t}\vert < \delta$, the target distribution is changed to a new one $\mathbf{q}_{i}\to\mathbf{q}_{i+1}$. In this variant, there are no proper trials but the target distribution keep changing until a maximum number
of rounds is reached.

\hfill \break
\noindent \textbf{Sinusoidal distribution shift} [\textsc{MAB-$\sin$}]\\
The reward distribution changes over rounds, with the probability of each arm following a sine wave with a specific frequency $f_{k}$, phase $\lambda_{k}$ and amplitude $1$. At any given time $t$, the distribution is $\mathbf{p}_{t}=\{\sin(2\pi f_{k} t+\lambda_{k})\text{  for }k=1\ldots K\}$.

\hfill \break
\noindent \textbf{Partial sinusoidal distribution shift} [\textsc{MAB-$\sin$P}]\\
Identical to the sinusoidal distribution shift, but a half of the arms change sinusoidally while the other half is always kept at a constant value. The distribution is not normalized.

\subsection{Evolution search}
The optimization of the hyper-parameters was performed using the Covariance Matrix Adaptation evolutionary strategy algorithm (CMA-ES) \cite{igel2007}.
The search was run with a population of $256$ individuals for $80$ generations. Each individual was endowed with a genome, corresponding to a vector of 22 parameters of the model.
The fitness function of the evolution was defined as the average reward obtained by an individual over 3 different non-stationary bandit environments, each for $K=\{40, 200\}$, and all averaged over 2 iterations.
The results are summarized below in figure \ref{fig:evolution}.

\begin{figure}[H]
    \centering
    \includegraphics[width=1.0\textwidth]{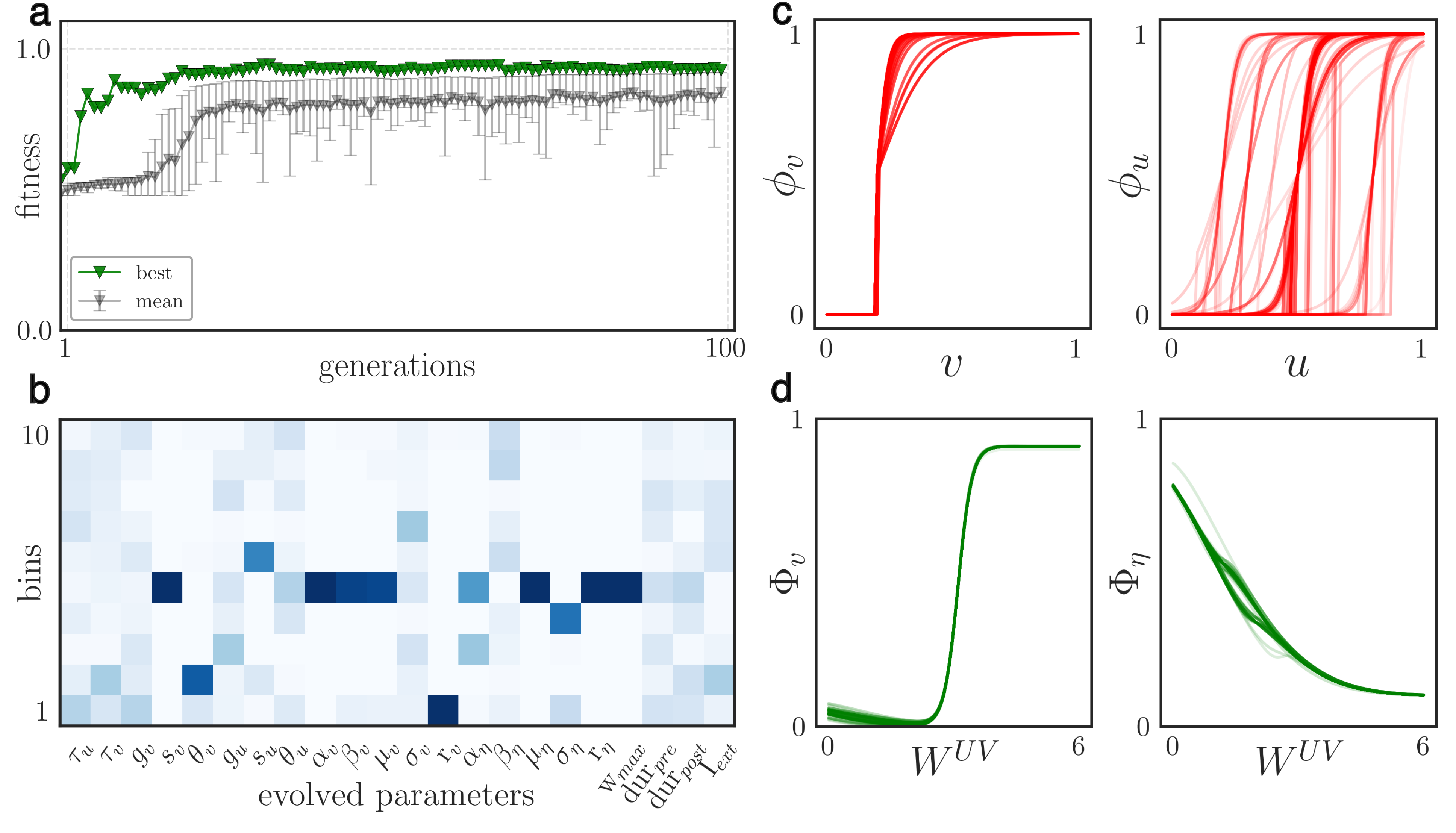}
    \caption{\textsc{Evolution results} - \textbf{a}: \textit{top fitness, mean and standard deviation (as 16-84 percentile) of the population over generations.} - \textbf{b}: \textit{heatmap of the evolved parameters (rows) as histogram bins (y-axis) calculated from the 50 percentile of the
    population of the last generation; higher density is in dark blue} - \textbf{c-d}: \textit{neural response functions [d] and Gaussian-sigmoid [c] of the top-half of the population, the color intensity is proportional to the fitness.}}
    \label{fig:evolution}
\end{figure}

\noindent The evolution progress, plot \textbf{\ref{fig:evolution}a}, showed a steady improvement over the generations, with the fitness hitting a plateau once the total reward could not get any closer to the optimum $\approx 1$ due to randomness and the finite horizon.

The evolved parameters (genome) are displayed as a frequency heatmap in \textbf{\ref{fig:evolution}b}, which was obtained from the individuals whose fitness was equal or above the median. The most relevant parameters resulted to be those concerning directly neuronal activity and learning.

More in details, the two Gaussian-sigmoid functions $\Phi_{v}, \Phi_{\eta}$, which are respectively the option value and learning rate functions, are shown in \textbf{\ref{fig:evolution}c}.
On the one hand, the learning rate function $\Phi_{\eta}$ is characterized by a decreasing curve, associated with the dominance of the right-side of the Gaussian, and a completely ignored sigmoid component.
A possible interpretation for this shape is to ensure a high learning speed when the value options are more uncertain (weak weights), and low otherwise; thus preventing overshooting and oscillations in the weight updates.
This adaptive behaviour is in line with known neuronal dynamics such as homeostatic plasticity, which works towards a stabilization of synapses, for instance through synaptic scaling and proportional updates \cite{citriSynapticPlasticityMultiple2008}.
A variable learning rate is an important feature of several plasticity rules, from the more biologically plausible like the Oja rule \cite{ojaOjaLearningRule2008} to deep learning optimizers like Adam \cite{kingmaAdamMethodStochastic2017}.

On the other hand, the option value function $\Phi_{v}$ follows instead a steep sigmoid curve.
This is consistent with the idea that the input of population $U$ to population $V$ is weighted maximally for high option values (strong synapses), whereas for weaker estimates the contributions are low or close to zero, allowing for more exploration.
Interestingly, a common feature seemed to be a slight concavity after zero, a slim influence of the Gaussian component, which might be interpreted as a sort of test for newly formed synapses. However, the size of this effect is not large.

The neural response functions are shown in \textbf{\ref{fig:evolution}d}. Both population evolved to have a similar shape, a sharp sigmoid with a clear threshold, with population \textit{U} having a more variable distribution.
The form is characterized by not allowing for a fine-grained linear reponse but rather an high-pass filter, with activity occurring only after strong excitation.
This firing behaviour is reminiscent of coincidence detector neurons, which are sometimes referred to as class III neurons with respect to their f-I curve \cite{ratteImpactNeuronalProperties2013}.

\subsection{Environment variants and number of arms}

The NSA model has been tested and compared with the other algorithms: Thompson Sampling (TS), $\epsilon$-Greedy, and UCB. The benchmark were the four different variants of the MAB problem listed above \ref{sec:envs}, with a variable number of arms ranging from 5 to 1000.
The results are reported in table \ref{tab:results}.
Overall, our NSA model displays a solid performance over all environments, most of the time being equally good or better than the other algorithms.
Interestingly, a large numbers of arms ($K$) did not present a significant challenge, as the model effectively adapted to the different environments and reward distributions.
However, this is in part due to the randomness in the assignment of arm probabilities, and the statistics of the quantity of high-reward arms as their number increases. Nonetheless, given the non-stationarity of the reward distribution it is still a non trivial task to re-calibrate to new distributions.


\begin{table}[ht]
\small
\centering
\begin{tabular}{c l c c c c c c}
\toprule
& \textbf{K} & \rotatebox[origin=c]{0}{$5$} & \rotatebox[origin=c]{0}{$10$} & \rotatebox[origin=c]{0}{$50$} & \rotatebox[origin=c]{0}{$100$} & \rotatebox[origin=c]{0}{$200$} & \rotatebox[origin=c]{0}{$1000$} \\
\midrule

\multirow{4}{*}{\rotatebox[origin=c]{90}{\textbf{MAB-P}}}
& TS & $\mathbf{0.03(8)}$ & $\mathbf{0.02(6)}$ & $\mathbf{0.02(7)}$ & $\mathbf{0.04(5)}$ & $\mathbf{0.02(3)}$ & $0.16(2)$ \\
& $\epsilon$-Greedy & $0.05(14)$ & $0.07(5)$ & $0.08(7)$ & $0.15(6)$ & $0.08(2)$ & $0.10(4)$ \\
& UCB & $0.05(15)$ & $0.05(8)$ & $0.19(6)$ & $0.33(3)$ & $0.39(2)$ & $0.54(3)$ \\
& \textit{NSA} & $\mathit{0.08(13)}$ & $\mathit{0.07(11)}$ & $\mathit{0.07(14)}$ & $\mathit{0.07(8)}$ & $\mathit{0.09(9)}$ & $\mathbf{0.07(8)}$ \\
\midrule

\multirow{4}{*}{\rotatebox[origin=c]{90}{\textbf{MAB-D}}}
& TS & $\mathbf{0.03(6)}$ & $\mathbf{0.08(13)}$ & $0.16(6)$ & $\mathbf{0.19(3)}$ & $0.28(7)$ & $0.34(3)$ \\
& $\epsilon$-Greedy & $0.04(7)$ & $0.14(13)$ & $0.22(5)$ & $0.19(8)$ & $0.26(7)$ & $0.16(4)$ \\
& UCB & $0.05(6)$ & $0.09(13)$ & $0.21(3)$ & $0.36(4)$ & $0.40(3)$ & $0.49(2)$ \\
& \textit{NSA} & $\mathit{0.13(10)}$ & $\mathit{0.15(16)}$ & $\mathbf{0.05(6)}$ & $\mathit{0.21(5)}$ & $\mathbf{0.26(7)}$ & $\mathbf{0.12(7)}$ \\
\midrule

\multirow{4}{*}{\rotatebox[origin=c]{90}{\textbf{MAB-$\sin$}}}
& TS & $0.21(22)$ & $0.22(16)$ & $0.07(5)$ & $0.10(5)$ & $\mathbf{0.06(4)}$ & $0.21(5)$ \\
& $\epsilon$-Greedy & $0.21(21)$ & $0.18(10)$ & $0.12(5)$ & $0.12(6)$ & $0.10(4)$ & $0.10(1)$ \\
& UCB & $0.03(4)$ & $0.05(3)$ & $0.17(4)$ & $0.23(1)$ & $0.33(3)$ & $0.49(3)$ \\
& \textit{NSA} & $\mathbf{0.00(3)}$ & $\mathbf{0.02(4)}$ & $\mathbf{0.05(3)}$ & $\mathbf{0.06(4)}$ & $\mathit{0.08(1)}$ & $\mathbf{0.05(4)}$ \\
\midrule

\multirow{4}{*}{\rotatebox[origin=c]{90}{\textbf{MAB-$\sin$P}}}
& TS & $0.19(21)$ & $0.43(19)$ & $0.17(10)$ & $0.11(6)$ & $0.09(6)$ & $0.19(6)$ \\
& $\epsilon$-Greedy & $0.24(26)$ & $0.43(10)$ & $0.24(10)$ & $0.14(5)$ & $0.15(6)$ & $0.14(2)$ \\
& UCB & $0.00(6)$ & $0.26(17)$ & $0.18(6)$ & $0.29(3)$ & $0.34(1)$ & $0.52(3)$ \\
& \textit{NSA} & $\mathbf{0.00(9)}$ & $\mathbf{0.23(17)}$ & $\mathbf{0.14(7)}$ & $\mathbf{0.08(8)}$ & $\mathbf{0.06(4)}$ & $\mathbf{0.09(7)}$ \\\\
\bottomrule
\end{tabular}
\caption{\textsc{Table of performance} — \textit{From top to bottom: results for MAB-P, MAB-D, MAB-$\sin$, and MAB-$\sin$P, for different numbers $K$ of arms. Each cell shows average regret and standard deviation (in parentheses), computed over 2 trials of 2000 rounds, averaged over 5 simulations.}}
\label{tab:results}
\end{table}

\subsection{Analysis of dynamics and robustness}

\subsubsection{Entropy analysis}\label{sec:entropy}
\begin{figure}[H]
    \centering
    \includegraphics[width=1.0\textwidth]{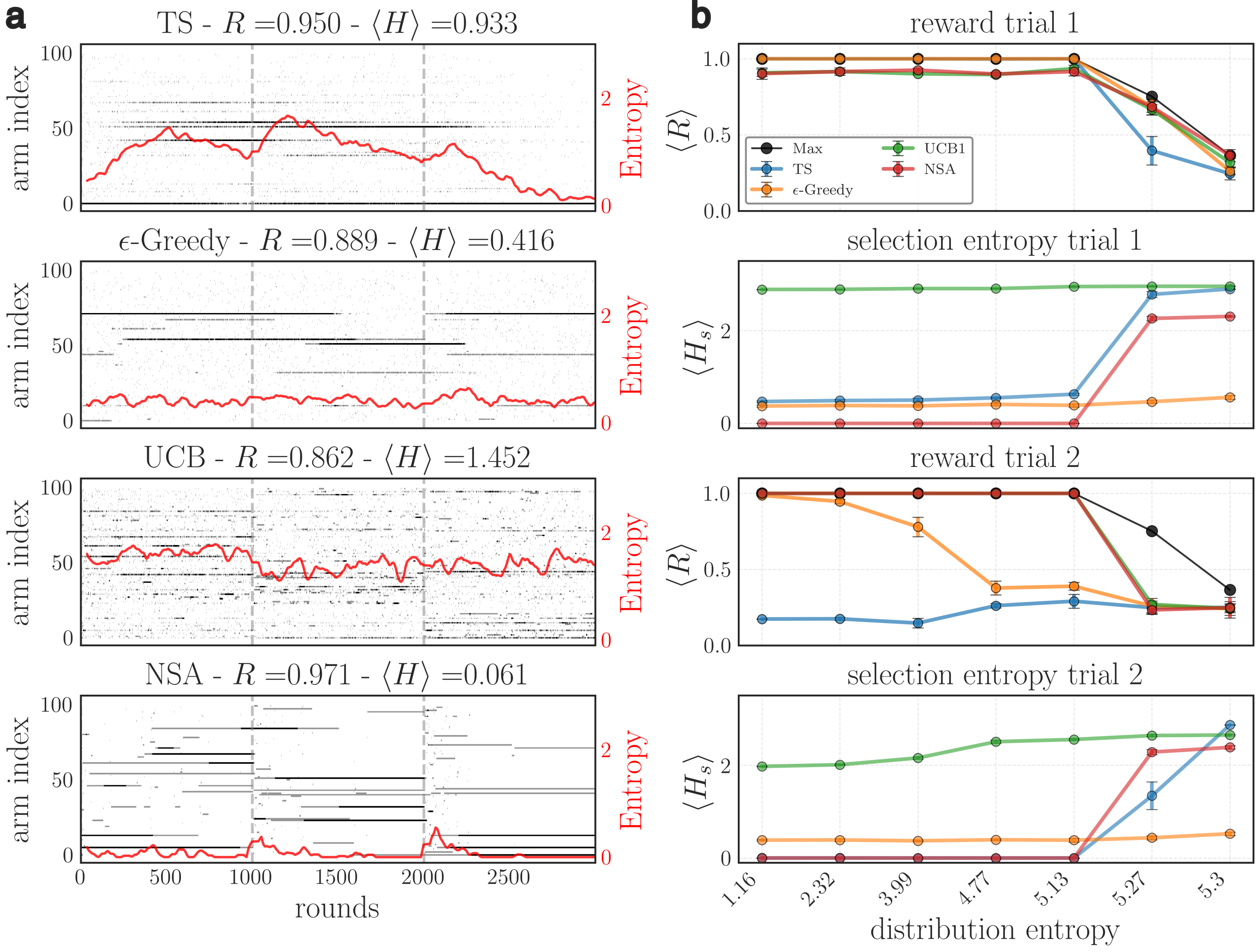}
    \caption{\textsc{Decision-making dynamics for different models} - \textbf{a}: \textit{Each plot display the results from one model. The raster plots (black dots) show the arms selected at each round.
The red lines represent the entropy level, calculated from the distribution of selections over the preceeding 20 rounds, smoothed with a 30-steps moving average. In the plot titles, the total reward and average entropy (H) over all trials are also reported.}
- \textbf{b}: \textit{rows 1-3 display the average reward $\langle R\rangle$ for trial 1 and 2 obtained by each model for increasing levels of distribution entropy (in nats) in the reward distribution. Rows 2-4 display the average entropy of the selections $\langle H_{s}\rangle$ for the first and second trial of the simulation, each with 2000 rounds.}}
    \label{fig:entropy_fig1}
\end{figure}

\noindent For a better understanding of the qualitative differences between the models, we analyzed the progress over the rounds by tracking the selected arms in a simple piecewise stationary distribution environment.
The simulation was run for $3$ trials with $2000$ rounds each and then averaged over $5$ iterations.
Furthermore, in order to quantify the variability of the decision policy at a given time and highlight the particularity of each decision-making behavior, we calculated the entropy of the probability distribution $p$ of the chosen arms, calculated over a window of 20 rounds, as $H=-\sum^{K}_{i} p_{i}\log(p_{i})$.
The unit of entropy is in nats, and it ranges from $0$ (no uncertainty) to $\log_{e}(K)$ (maximum uncertainty).
In Figure \ref{fig:entropy_fig1}-\textbf{a}, the raster plot of the selected arms is plotted for each model together with its level of entropy. The distribution of the probability of reward on the arms has an average of $H=2.02$.

As expected, the shape of the entropy curve expresses the inherent strategy adopted by each model.
In particular, the UCB algorithm showed the highest variability, marked by persistent exploratory behavior throughout the trials despite converging to reward options. Thompson Sampling was able to reach most solutions, although it had difficulty adapting to new reward distributions
that led to high entropy levels.
$\epsilon$-Greedy also showed a good performance quite reliably, with the greedy strategy ensuring low entropy for most rounds.
Similar behaviour was observed for NSA, which was able to reach the optimal policy and maintain it over time, with entropy peaking mostly at the beginning of the trials and being, on average, the lowest among all models.
Indeed, the dynamics of NSA makes it particularly suited for the task of non-stationary MAB, as it is able to quickly adapt to new reward distributions and firmly maintain a greedy policy.

\noindent Then, we sought to investigate the robustness of NSA by targeting the capacity to endure increasing levels of entropy in the reward distribution, defined as $-\sum^{N}_{i} p_{i} \log(p_{i})$ and calculated in nats.
In particular, this analysis tracked performance, cumulative reward, as well as selection entropy $H_{s}$, defined as the entropy of the distribution of selected options within a sliding window of $20$ rounds.

The simulation was carried out in a piecewise stationary environment with $K=200$ in two trials averaged over 5 independent runs, and it is shown in Figure \textbf{\ref{fig:entropy_fig1}b}.
The distributions were chosen such that they have only one strongly rewarding arm in order to highlight the models' ability to find it.
For more details on the distribution, see the Appendix \ref{sec:appendix_entropy}.
In the top two plots, the average reward and regret obtained by each model is shown against the entropy of the reward distribution for the two trials.

The results reported how all models are capable of robust performance in the first trial even in the presence of high uncertainty.
In the second trial, $\epsilon$-Greedy and Thompson Sampling suffered the increasing difficulty of switching arms, probably due to their conservative approaches. However, this challenge affected UCB and NSA only with higher entropy levels, recognizing their adaptability.

Another perspective on this analysis was given by the two bottom plots, which showed the average entropy over the trials.
Overall, there was an unsuprising trend of increasing selection entropy with the entropy of the reward distribution.
However, striking is the exception of Epsilon-Greedy, which still maintained a constant level throughout.
UCB displayed the highest average values, while Thompson Sampling followed with some delay.
On the other hand, NSA displays a more abrupt change, going from a state of very low to high variability, a sign of solid exploratory behavior.




\section{Discussion}

The ability to make decisions under uncertainty is a fundamental aspect of cognition. A well-established framework for studying this capacity is the multi-armed bandit problem (MAB), which has been widely explored and extended across multiple domains \cite{suttonReinforcementLearningProblem1998, liu2024a}.

In behavioral experiments, humans demonstrate remarkable adaptability in such settings, integrating environmental uncertainty, generalizing across trials, and dynamically adjusting their learning rates. These behaviors reflect a diversity of cognitive strategies \cite{steyversBayesianAnalysisHuman2009a}. Although Bayesian approaches often capture human behavior well \cite{behrens2007, schulzFindingStructureMultiarmed2020, zhangForgetfulBayesMyopic2013}, they are challenging to map directly onto biologically realistic neural dynamics. Despite the existence of many algorithms with strong theoretical guarantees, most lack biological plausibility, particularly in their architectural assumptions and learning and choice mechanisms.

In this work, our aim was to design a minimal, biologically inspired architecture able to solve non-stationary MAB tasks.
Specifically, we proposed a simple architecture composed of two interacting and plastic populations of rate-based neurons and producing choices through agreement, called Neural Selection Agreement model (NSA).
We evaluated it on four variants of the MAB problem, each differing in how reward probabilities evolved over time and in a wide range of arm counts, from a few options to over a thousand. For comparison, we also tested three standard algorithms.

Our results show the model's ability to adapt to changing reward distributions and quickly recover performance over time.
It reliably tracked reward-optimal options and sustained effective decision policies, matching the performance of established methods such as Thompson Sampling, $\epsilon$-Greedy, and Upper Confidence Bound (UCB).

To better understand the behavior of the system, we analyzed its responses at varying levels of reward distribution entropy.
In low-uncertainty settings, NSA quickly identified the rewarding option and adopted a greedy policy, similar to Thompson Sampling.
In contrast, UCB maintained a higher degree of exploration. As uncertainty increased, the model exhibited a higher option entropy in its decisions, transitioning to a more exploratory strategy similar to UCB.
Although this change modestly affected the NSA's ability to switch arms in highly volatile environments, overall performance remained robust.

The strengths of the our model can be traced in both the architecture and the learning paradigm, whose hyperparameters were optimized through an evolutionary process. Interestingly, the values found converged to solutions that can be mapped to plausible synaptic mechanisms.
On the one hand, neural dynamics, which rely on plastic connections and a consensus-like selection process.
Particularly important was the choice of modulating the afferent connections to the value population $V$ according to a nonlinear function dependent on the synaptic weight itself. In so doing, it was possible to implicitly evolve an effective option value policy for the trade-off between exploration and exploitation.

The neural response functions that emerged were characterized by a steep sigmoidal shape, which can be related to the saturation of the neural response once a certain threshold is crossed, a feature observed in the biological network as class III neurons, in addition to being a common choice for artificial ones \cite{ratteImpactNeuronalProperties2013, ockerFlexibleNeuralConnectivity2020, apicellaSurveyModernTrainable2021}.

On the other hand, learning was structured as a nonassociative plasticity rule based on the reward. Similarly to before, a non-linear function of synaptic weights played a critical role, specifically in defining the synapse-specific learning rate \cite{larsenSynapsetypespecificPlasticityLocal2015}.
Furthermore, the shape evolved of the learning rate function was inversely proportional to the synaptic weight, which can be related to the availability of resources in the synapse and its state, including size \cite{bartolHippocampalSpineHead2015, arielIntrinsicVariabilityPv2012}.

An additional consideration is the inspiration from the functional role of the orbitofrontal cortex (OFC) and anterior cingulate cortex (ACC), two important pre-frontal regions known to be involved in decision-making processes \cite{kennerleyDecisionMakingReward2011a, khamassiChapter22Medial2013}.

In our work, we have explored ways in which an option value can be formed according to recent reward history and connection weights, updating the option representations.
The OFC is known to represent different options, updating their values based on history and rewards \cite{lukChoiceCodingFrontal2013, kennerleyDecisionMakingReward2011a, klein-flugge2013}.

Next, the NSA model relies on some inductive biases, the shape of the Gaussian-sigmoid function, affecting the neural activity and the weight-dependent learning rate.
These biases may be considered to implicitly encode a policy that dictates how the two populations interact, how new information should be incorporated into the update of the weight value, and what option to choose next.
In fact, ACC has been associated with the evaluation of actions and the regulation of the balance between exploration and exploitation \cite{khamassiChapter22Medial2013, kolling2016}.

Additionally, the generation of an option selection results from a temporal interaction of the activity of the two neural population, before converging to a choices.
In a similar direction, it has been observed that the OFC transiently visits chosen and unchosen options before committing \cite{rich2016}.
Further, the dynamic interaction between the ACC and OFC has been linked to transient pre-stimulus activations, which bias decisions toward the most valuable option \cite{funahashiPrefrontalContributionDecisionMaking2017, marcosDeterminingMonkeyFree2016, balewskiValueDynamicsAffect2023}.

Despite the promising results, there are some limitations to the model. First, we considered the great level of abstraction in the neuronal details, as we considered simple point neurons with synapses modeled with relatively elementary functions, lacking the anatomical complexity of actual dendrites.
In particular, the NSA does not account for the presence of noise in neural dynamics, which is a well-known feature of biological neurons \cite{faisalNoiseNeuronsOther2012}.
Furthermore, the functional association with the pre-frontal cortical region is only moderate, although present.
On the computational side, since our interest lied in the biological plausibility and evolution of adaptive meta-learning solutions, we used only a few well established and relatively simple algorithms as a reference and did not take into account more advanced variants such as VDBE \cite{tokicAdaptiveEGreedyExploration2010, tokicValueDifferenceBasedExploration2011}, \textit{f-Discounted-Sliding-Window Thompson Sampling} (\textit{f-dsw TS}) \cite{cavenaghiNonStationaryMultiArmed2021}, and variants of $\epsilon$-Greedy \cite{qiForcedExplorationBandit2023}.

Future work could involve comparison with more sophisticated algorithms, the introduction of a larger architecture, and more realistic neural dynamics, such as spiking neurons \cite{nunesSpikingNeuralNetworks2022}.



\hfill \break
\vspace {0.2cm}

\noindent \textbf{Acknowledgements \& Statements}\\
The authors declare no competing interests.
\hfill \break
The code is publicly available and can be found at
\url{https://github.com/iKiru-hub/minBandit.git}.\\
This research was funded by the European Union’s Horizon 2020 research and innovation programme under the Marie Skłodowska-Curie grant agreement Nº 945371 and the University of Oslo. \\
The research presented in this paper has benefited from the Experimental Infrastructure for Exploration of Exascale Computing (eX3), which is financially supported by the Research Council of Norway under contract 270053.

\noindent Lastly, special thanks to Kosio Beshkov and Marianne Fyhn for inputs and feedback.

\vspace {0.3cm}

\bibliography{mkb_bibliography}

\newpage

\section{Appendix}\label{sec:appendix}

\subsection{Neural response function}

The activation functions applied to the two neuronal population are defined as a step-function composed with a generalized sigmoid as follows:

\begin{equation}
f(x; g, o, \theta) =
\begin{cases}
    \left[1 + e^{-g(x - o)}\right]^{-1} & \text{ if } \left[1 + e^{-g(x - o)}\right]^{-1} > \theta \\
0 & \text{ otherwise}
\end{cases}
\end{equation}

Where:
\begin{itemize}
\item $x$ is the neuron pre-activation value
\item $g$ is the gain
\item $o$ is the offset
\item $\theta$ is the threshold
\end{itemize}
\noindent Each population has its own set of parameters, which are optimized through evolutionary search.

\subsection{Gaussian-sigmoid function}

\noindent The function $\Phi_{\cdot}$ is defined by combining a generalized version of the sigmoid, namely with a gain $\beta \neq 1$ and offset $\alpha\neq 0$, and a Gaussian with mean $\mu$ and variance $\sigma^{2}$. Their contributions are weighted by as $r$ and $1-r$ ($r\in(0,1)$) respectively.

\begin{equation*}
    \Phi_v(x) = r\left(1 + \exp^{-\beta(x-\alpha)}\right)^{-1} + (1-r)\exp\left(-\frac{(x-\mu)^2}{2\sigma^2}\right)
\end{equation*}

\noindent The motivation behind this choice is to express a function that possesses a bounded region (depending on $\mu,\,\sigma$) at a high/low peak (depeding on the value of $\gamma_{2}$), and a continuous transition to a constant value (depending on the steepness of the sigmoid $\beta$, shift
$\alpha$, and intensity $\gamma_{1}$).

\begin{figure}[H]
    \centering
    \includegraphics[width=0.8\textwidth]{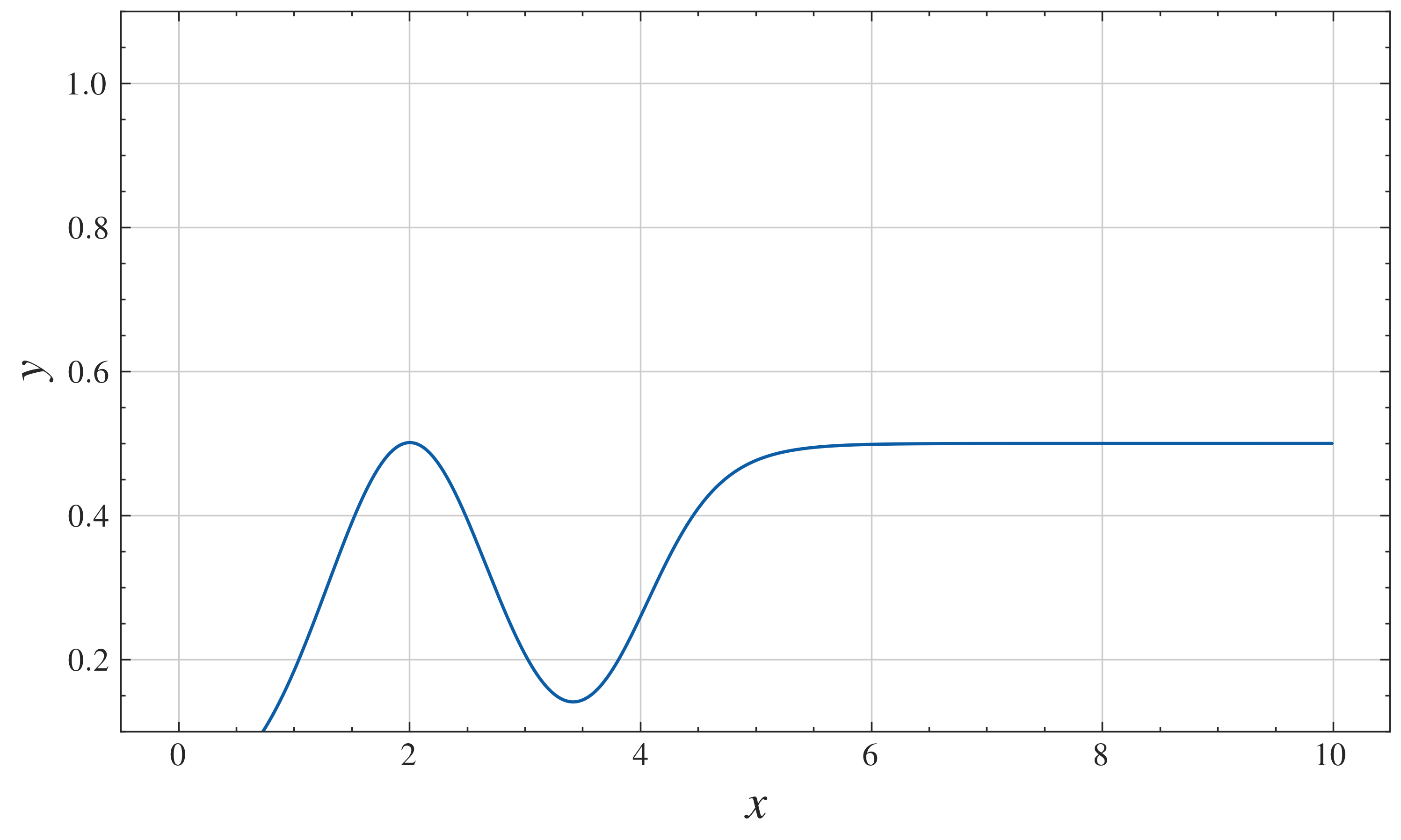}
    \caption{\textsc{Activation function $\Phi_{v}$} - \textit{Parameters $\beta=10$, $\alpha=1$, $\mu=1$, $\sigma=1$, and $r=0.5$.}}
    \label{fig:gau_sigm}
\end{figure}

\subsection{Evolution search}
The optimization was carried out over several parameters concerning the model architecture and dynamics:

\noindent \textbf{Network parameters}
\begin{itemize}
    \item $\tau_{u}$: time constant of population $U$
    \item $\tau_{v}$: time constant of population $V$
    \item $g_{u}$: gain of the neural response function of population $U$
    \item $g_{v}$: gain of the neural response function of population $V$
    \item $o_{u}$: offset of the neural response function of population $U$
    \item $o_{v}$: offset of the neural response function of population $V$
    \item $\theta_{u}$: threshold of the neural response function of population $U$
    \item $\theta_{v}$: threshold of the neural response function of population $V$
    \item $W^{+}$: maximal weight value for the weights $\textbf{W}^{UV}$
\end{itemize}

\noindent \textbf{Option value function parameters}
\begin{itemize}
    \item $\beta_{v}$: steepness of the sigmoid
    \item $\alpha_{v}$: shift of the sigmoid
    \item $\mu_{v}$: mean of the Gaussian
    \item $\sigma_{v}$: variance of the Gaussian
    \item $r_{v}$: weight of the sigmoid
\end{itemize}

\noindent \textbf{Learning rate function parameters}
\begin{itemize}
    \item $\beta_{\eta}$: steepness of the sigmoid
    \item $\alpha_{\eta}$: shift of the sigmoid
    \item $\mu_{\eta}$: mean of the Gaussian
    \item $\sigma_{\eta}$: variance of the Gaussian
    \item $r_{\eta}$: weight of the sigmoid
\end{itemize}

\noindent Each individual has been evaluated over environment the following environments:

\begin{itemize}
    \item \textsc{MAB-0}: average reward distribution entropy $\langle H\rangle=2.05$
    \item \textsc{KAB-$\sin$P}: average reward distribution entropy $\langle H\rangle=2.1$, given $K$ arm frequencies $f_{k}$ as an equally spaced set $\{0.1\ldots i\ldots 0.4\}$, phases $\lambda_{k}$ drawn from an uniform $\sim \mathcal{U}(0, 2\pi)$, and half of the arms have been set to constant values drawn from
        another uniform $\sim \mathcal{U}(0.1, 0.7)$; the final reward distribution was not normalized.
\end{itemize}

\noindent The number of arms was $K=10$ and $150$, and lasted for $2$ trials with $2000$ rounds each.
The final fitenss was the average over $2$ iterations.

\hfill \break
The optimization has been implemented in Python using the \texttt{DEAP} library, and the algorithm used was the \texttt{CMA-ES} algorithm. The optimization involved $40$ generations with a population size of $256$ individuals. The mutation rate was set to $0.5$ with a sigma of $0.8$, the cross-over rate was set to $0.4$.
The run were carried out on a 256-core AMD EPYC 7763 with 2TB of RAM.

\subsubsection{Genome distribution}
Following the evolution search, it is taken the distribution of parameters over the top-scoring half of the population, corresponding to 128 individuals.

\begin{figure}[H]
    \centering
    \includegraphics[width=1.\textwidth]{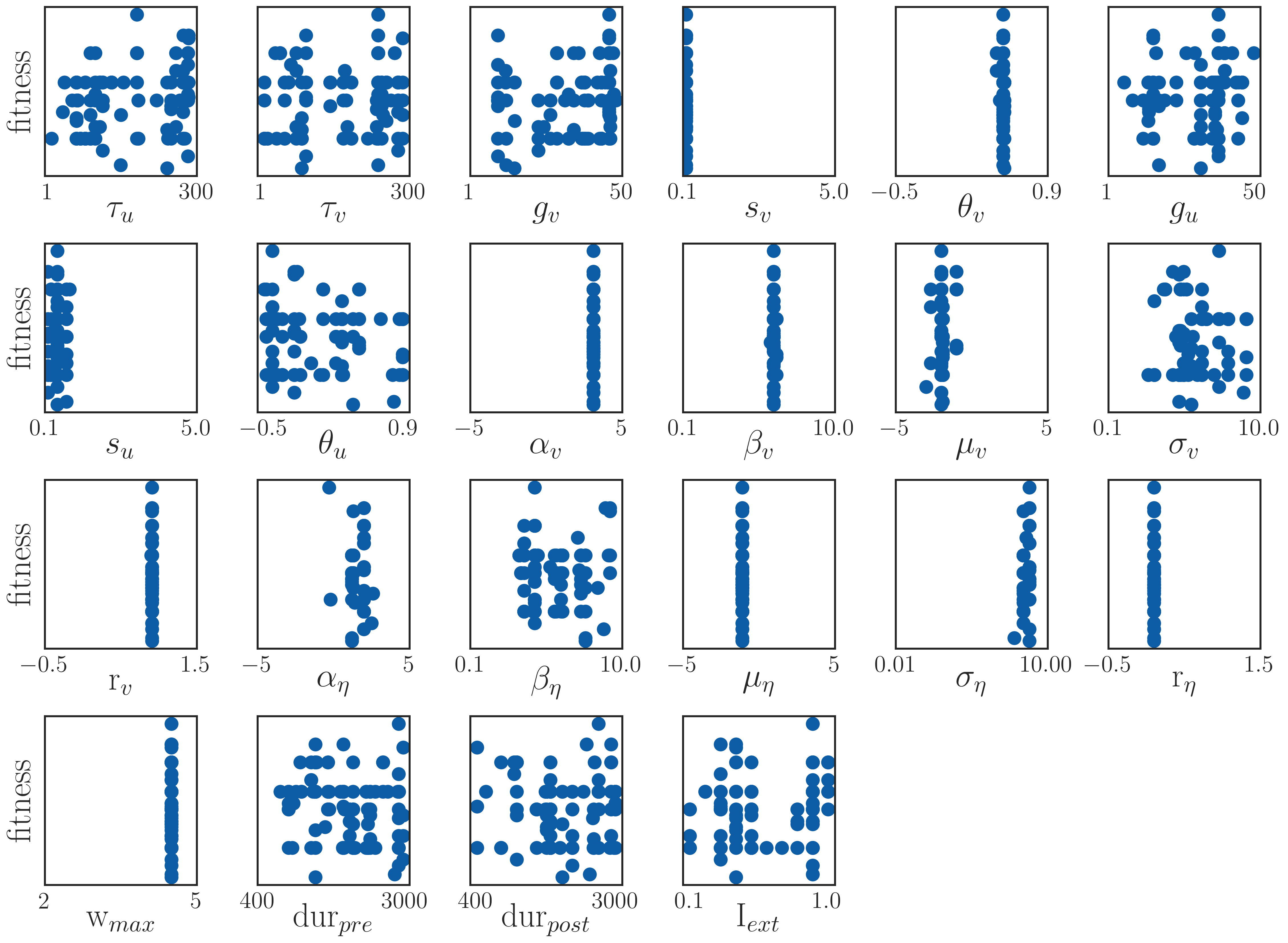}
    \caption{\textsc{Genome distribution} - \textit{each parameter is plotted against the fitness score.}}
    \label{fig:evo_tabplot}
\end{figure}

\noindent Figure \ref{fig:evo_tabplot} reveals those parameters shared among the models with the highest fitness score, and those that are more variable.
The most stable parameters are, predicatbly, those involved in directly shaping the neural activity (neural activation functions) and learning policy (Gaussian sigmoid).

\subsection{Reward distribution entropy}\label{sec:appendix_entropy}

\noindent The calculation of a set of $N$ reward probability distribution $\mathbf{p}_{i}\text{  for  } i\ldots N$ for $K$ values with a progressively decreasing levels of entropy $\mathbf{h}_{i}\text{  for  } i\ldots N$ has been obtained by the algorithm below \ref{alg:entr_alg}.

\begin{algorithm}[H]
\caption{Reward Probability Distribution Generation}
\label{alg:reward_distribution}
\SetAlgoLined
\KwIn{Number of distributions $N$, dimension $K$}
\KwOut{Set of probability distributions ${\mathbf{p}_i}$ with decreasing entropy}
\SetKwComment{Comment}{// }{ }
\textbf{Initial Setup:}
Define set $B = \{17, 15, 12, 8, 4, 1.5, 0.5\}$; \\
\For{$i \gets 1$ to $N$}{
    $\mathbf{z} \gets \text{RandomVector}\sim \mathcal{U}(0,0.5)^K$;\\
    $j \gets \text{RandomIndex}(K)$;\\
    $\mathbf{z}_j \gets 1$;\\
    $\beta_i \gets \text{Sample index=} i \text{ from }(B)$ \Comment*[r]{Sample temperature from $B$}

$\mathbf{p}_i \gets \frac{\exp(\beta_i \mathbf{z})}{\sum_j \exp(\beta_i \mathbf{z}_j)}$ \Comment*[r]{Softmax with temperature}
}
\Return ${\mathbf{p}_i}$
\end{algorithm}\label{alg:entr_alg}

\subsection{Weight update dynamics}
\begin{figure}[H]
    \centering
    \includegraphics[width=1.0\textwidth]{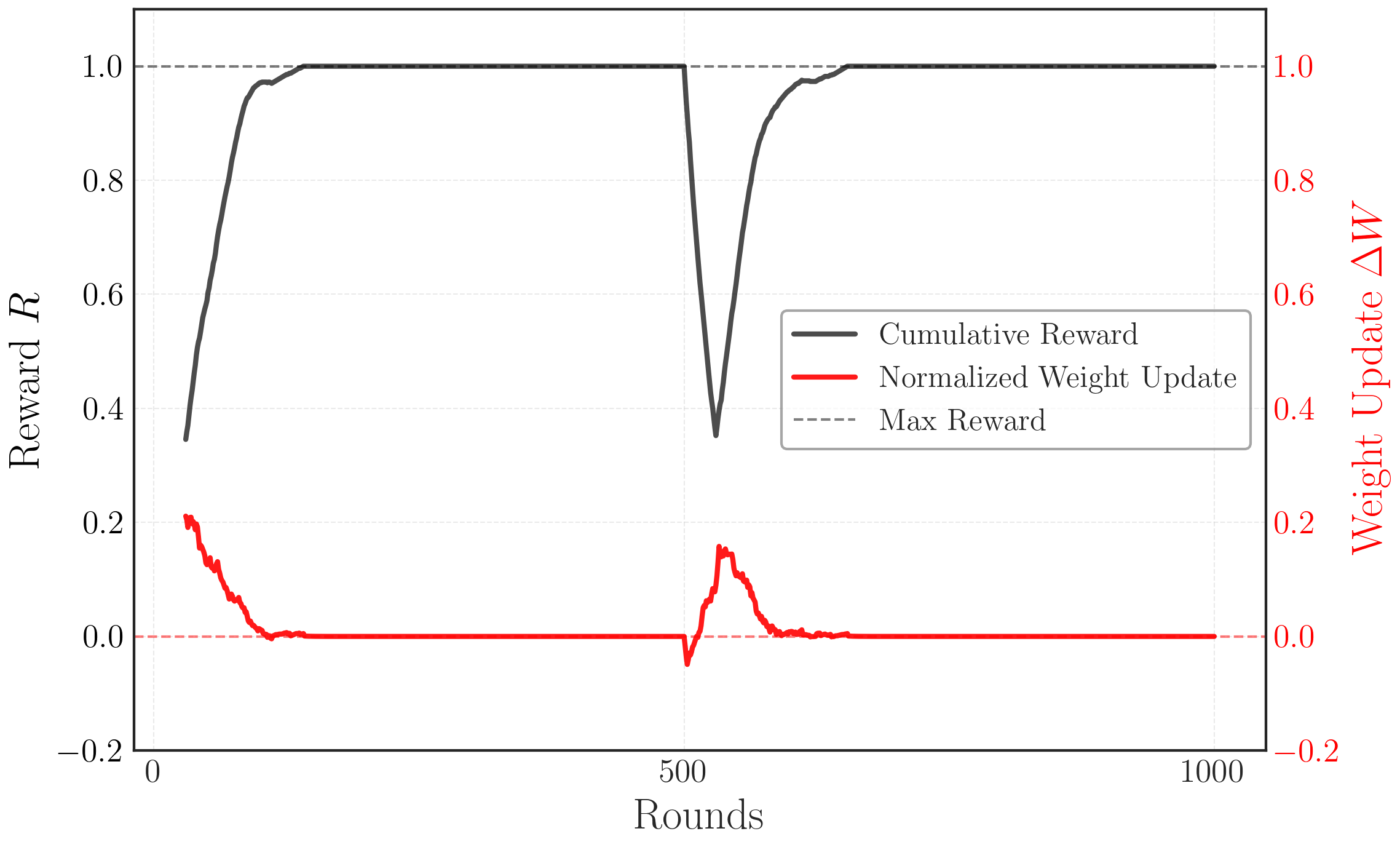}
    \caption{\textsc{Weight update developement for the model} \textit{The plot displays the weight update quantity $\Delta W_{k}^{UV}$ for each round (blue line), smoothed as a 20-steps moving average.
    It is also reported the average reward in a window of 30 rounds (orange line). The results have been obtained averaging over 30 iterations.}}
    \label{fig:rew_update}
\end{figure}

\noindent We also analyzed the weight update dynamics of the model over the rounds.
In figure \ref{fig:rew_update}, we plotted the evolution of the total weight $\Delta W^{UV}$ over time, averaged over 20 simulations, and smoothed over 30 rounds.
The results show that the model can quickly adapt to new reward distributions. It is also able to maintain the optimal policy over time, with the weights remaining approximately stable.
The quantity of updates $\Delta W_{k}^{UV}$, which in each round is applied to one connection $k$, changes sign according to the collected reward, with its magnitude higher at the beginning of the trials.
Initially, the sign is mostly positive (potentiation) since the weights start at zero, and after some uncertainty a consistently preferred arm emerges.
However, when the reward distribution switches, a regular series of suboptimal choices with respect to the new distribution is made, leading to zero reward.
This causes an accumulation of negative sign weight updates (depression), eventually causing the value of the preferred arm to drop. In the meantime, other options are probed until another sequence of choices converges to another arm, promoted by a trail of positive weight updates.

This behaviour is consistent with the low entropy levels observed in the previous analysis.

\end{document}